\documentclass[journal]{IEEEtran}

\usepackage{graphicx}
\usepackage{subcaption}
\ifCLASSINFOpdf
\else
\fi
\begin{document}
%
\title{Tokenized Model: A Blockchain-Empowered Decentralized Model Ownership Verification Platform}
%
%
%

\author{Yihao Li,
        Yanyi Lai,
        Tianchi Liao,
        Chuan Chen*,~\IEEEmembership{Member, IEEE,}
        and Zibin Zheng,~\IEEEmembership{Fellow, IEEE}
        
    

\thanks{Yihao Li, Yanyi Lai and Chuan Chen are with the \textit{Department
of School of Computer Science, Sun-Yat sen University, Guangdong, 510275, China. }
Tianchi Liao and Zibin Zheng are with the \textit{Department of School of Software Engineering, Sun-Yat sen University, Guangdong, 519000, China.}}
\thanks{Corresponding author E-mail: chenchuan@mail.sysu.edu.cn}
}

%
%

\markboth{Journal of \LaTeX\ Class Files,~Vol.~14, No.~8, August~2015}%
{Shell \MakeLowercase{\textit{et al.}}: Bare Demo of IEEEtran.cls for IEEE Journals}
%



\maketitle

\begin{abstract}
With the development of practical deep learning models like generative AI, their excellent performance has brought huge economic value. For instance, ChatGPT has attracted more than 100 million users in three months. Since the model training requires a lot of data and computing power, a well-performing deep learning model is behind a huge effort and cost. Facing various model attacks, unauthorized use and abuse from the network that threaten the interests of model owners, in addition to considering legal and other administrative measures, it is equally important to protect the model’s copyright from the technical means. By using the model watermarking technology, we point out the possibility of  building a unified platform for model ownership verification. Given the application history of blockchain in copyright verification and the drawbacks of a centralized third-party, this paper considers combining model watermarking technology and blockchain to build a unified model copyright protection platform. By a new solution we called Tokenized Model, it protects the model’s copyright by reliable ownership record and verification mechanism. It also promotes the financial value of model by constructing the model’s transaction process and contribution shares of a model. In the typical case study, we also study the various performance under usual scenario to verify the effectiveness of this platform.
\end{abstract}


%
\IEEEpeerreviewmaketitle

\section{Introduction}
%
%
%
%



 

\subsection{Background}

\par In recent years, Artificial Intelligence Generated Content (AIGC) has experienced rapid development and demonstrated strong performance in various domains such as text, speech, and image generation. Applications like ChatGPT and Stable Diffusion have attracted considerable attention since released. Given the powerful capabilities of AIGC models, they can be employed to generate diverse types of content and generate revenue. Therefore, AIGC models hold significant commercial value in many application areas such as advertising, gaming, news, and artistic creation. Model owners who legitimately operate AIGC models stand to gain substantial financial returns. Correspondingly, the training of generative models typically requires substantial computational resources, often accessible only to large enterprises or institutions capable of bearing the costs. This circumstance increases the likelihood of model stealing. Malicious attackers can employ various techniques to steal a well-trained generative model at a low cost, exploiting it for their own profit, thereby infringing upon the rights of model owners and giving rise to copyright disputes. Hu \emph{et al.} \cite{hu2021model} have explored the feasibility of stealing generative models, providing evidence that the aforementioned concerns are not unfounded. Without a reliable model authentication mechanism, the model cannot serve as an asset for financial institutions or investors, thereby restricting the circulation and transaction of the model, making it difficult to conduct financial applications.

\par In order to address the issue of model copyright, model watermarking technology has become increasingly important. Model watermarking technology is a technique that embeds specific identifying information into the model. By incorporating watermarks into the model, it can be uniquely identified, and the generated content can be tracked to determine its source and the model ownership. 

\begin{figure}[!t]
    \centering
    \includegraphics[width=0.45\textwidth]{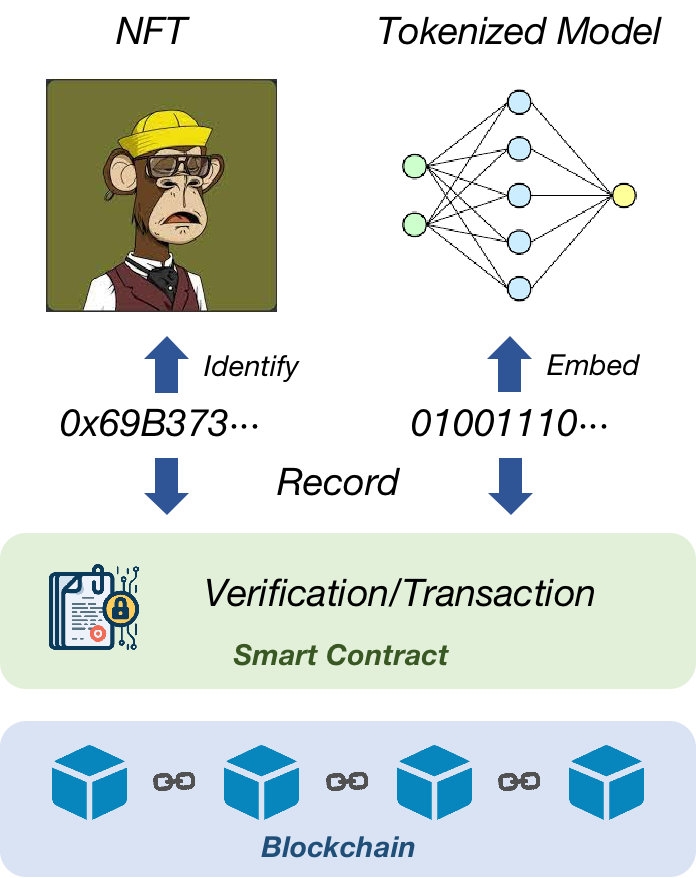} 
    \caption{Comparison between NFT and Tokenized Model.}
    \label{fig:NFT}
\end{figure}

\subsection{Motivation}
\par As described in the background, the copyright issues faced by AIGC have made the need for model authentication increasingly urgent. With the widespread application of generative models and the substantial amount of content they create, the protection of the model itself as a creative entity becomes crucial. The need for model authentication has become a practical necessity rather than merely a potential precautionary measure \cite{liang2023adversarial}.

\par Although existing model watermarking techniques can provide a certain level of security for models, in practical applications, the party conducting copyright authentication typically needs to know the embedded watermark information in advance for verification. The difficulty lies in providing proof without disclosing the watermark information. Revealing the embedded watermark information to the outside world can pose risks, as a malicious verifier can potentially retrain the model using the watermark information for model forgery, or exploit it as a tool to erase the original model's watermark for model stealing. Adi \emph{et al.} \cite{adi2018turning} have expressed this concern and have proposed blockchain as one possible solution.

\par To address the aforementioned issues, a unified model copyright authentication platform is needed. This platform would be responsible for watermark management and model copyright authentication. By directly querying the platform, one can have the copyright information without the knowledge of the embedded watermark with the reliable authentication process. By verifying the copyright permanently, the platform makes the model transaction possible, which can facilitate model circulation and broader applications. Centralized platforms are susceptible to fraud, opacity, single point of failure, making them unsuitable for serving as model copyright management platforms \cite{liang2020dual}. In contrast, blockchain technology possesses the characteristics of decentralization, immutability, and public transparency, offering significant advantages for applications involving the verification of copyright ownership.

\subsection{Contribution}
\par In this article, we propose a blockchain-empowered decentralized model ownership verification platform. The model are connected with its copyright information which stored on the blockchain like NFT, and thus we called this structure \textit{Tokenized Model}. (Figure \ref{fig:NFT} illustrates the comparison between NFT and Tokenized Model.) This platform enables model copyright verification, transactions, and supports multi-client scenarios of federated learning. The main contributions of this work are three-fold:
\begin{itemize}
    \item We propose a blockchain-empowered decentralized model ownership verification platform, where all model information recorded on the platform possesses the characteristics of immutability and non-repudiation, effectively protecting the model's copyright.
    \item We propose the tokenization of deep learning models, utilizing blockchain platforms and smart contracts for token distribution, management, verification, and transactions.
    \item Leveraging the blockchain platform, we facilitate the efficient circulation of models, making models tradable commodities similar to AIGC in terms of marketability.
\end{itemize}

\section{Related Works}
\par The application of blockchain to the storage and copyright verification of digital assets has a long history. Below we mainly introduce the relative work on blockchain concerned with copyright protection. 
\par Ethereum introduces programmable smart contracts into the blockchain to help create various decentralized applications. One of the most famous application is Non-Fungible Token (NFT) \cite{wang2021non}. Token is created through smart contract to help people record any digital assets on chain and make transaction based on a strict authentication mechanism. Fungible token only has the concept of quantity, but NFT has attributes that can uniquely determine itself like a paint or a music.  Based on these convinient, many digital assets such as paintings, music, videos, etc, can be managed by blockchain. Liu \emph{et al.} \cite{liu2021blockchain} designs a blockchain-based management platform for multimedia resources which point out the difficulty for a centralized system to verify the integrity of data and designs a multi-level access control mechanism. Adjei-Mensah \emph{et al.} \cite{adjei2021securing} introduces an internet database platform for music creation that utilizes blockchain technology to store music works and protect copyright information of music albums. By leveraging the decentralized and tamper-proof features of the Ethereum blockchain, this platform provides a secure and reliable solution for music creators. Chen \emph{et al.} \cite{chen2020dcdchain} designs an architecture for blockchain to protect and verify copyright of  traditional digital assets which mainly focus on building a reliable verification process.
\par The application of blockchain in copyright protection informs us that a reliable model ownership verification platform can be empowered by blockchain. As the model parameters grow larger and layers grow deeper, the practice of directly uploading original data on the chain like other digital assets is impractical for models. This article will explore how to tokenize deep learning models and give them properties like other NFTs, which means it can be verified and given liquidity. Before the architecture, we also need to be familiar with model watermarking technology. 
\par Watermark is often embedded into digital resources to protect copyright and protect from data leakage. The watermark can tell us the information of leakers and the other concerned with the copyright.  However, due to black box nature of deep learning models, traditional methods cannot insert watermarks into models without affecting their original functions. Therefore, model watermarking technology has attracted many researchers. The current model watermarking method integrates the watermark embedding process into the training process and avoids affecting the original function of the model through the training process by designing the specific loss function. Below, we introduce three main types of model watermarking:
\begin{itemize}
\item Backdoor watermarks \cite{peng2023you}: Originates from backdoor attacks \cite{guo2022overview}, where the attacker merges a designed trigger into the training dataset and identifies the samples containing the trigger as a specific label instead of their true labels during the model training process. Therefore, the model will generate malicious result for the trigger-contained data and tend to be normal facing normal data. In the model watermarking method, this trigger can be seen as a non-visible watermark and by observing the model’s output for the samples containing the trigger, we can know whether the model contains the watermark or not.
\item Parameter features \cite{shao2022fedtracker}: Inspired by traditional methods, where we hide the watermark into the parameters of the model. We can decode and get the watermark by a specific designed decoding process. As mentioned earlier, since it is not clear how changing model-specific parameters will affect the neural network function, some optimization or deep learning methods are still needed to assist the embedding process to ensure that the functionality of the model is not affected. The advantage of this approach is that the watermark is more stable and can hide information into models.
\item   Generated object \cite{fernandez2023stable}: Designed for generative models. The generative modules like VAE and GAN of  generative models can be trained so that the generators contain special watermarks that can be detected by decoders.  We can verify the origin of a generative content by detecting the existence of the watermark. 
\end{itemize}

\begin{figure*}[!t]
    \centering
    \includegraphics[width=\textwidth]{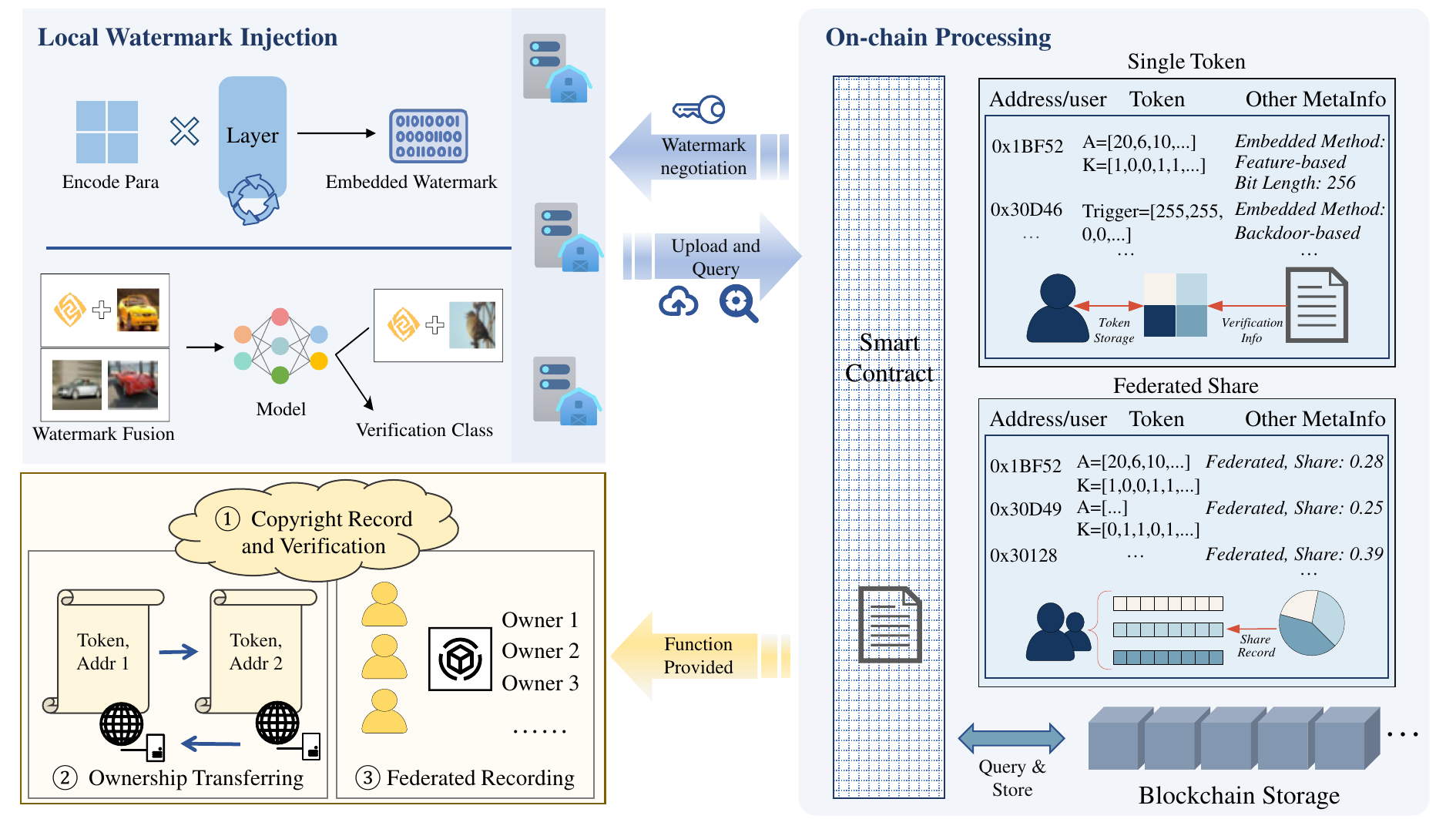} 
    \caption{Overall Framework of TokenizedModel. }
\end{figure*}

\section{Tokenized Model and Platform Architecture}

\par In this section, we will discuss the proposed overall architecture. We will refer to this framework as the \textit{TokenizedModel} in the following text. The overall procedure and functions of this architecture are shown in Figure 2. In this framework, our goal is to manage model watermarking through blockchain and provide more practical application services based on this basic function of copyright verification and recording. The entire process is divided into local watermark injection and on-chain processing  through smart contracts. In Figure 2, the part in blue is the basic process and in yellow the function that the system provides. We will first introduce these basic process and then the core functions of this copyright verification platform.

\subsection{Overall Architecture}
\subsubsection{Local Watermark Injection}
\par In previous works of model watermarking, whether by backdoor watermarking or by feature watermarking, clients will choose a specific  watermark themselves. However in \textit{TokenizedModel},  the client need to negotiate with the blockchain to get their own watermark. Suppose the client decides to use feature watermark, it will get a decoded matrix \textbf{A} and a watermark \textbf{K}. The negotiation process prevent the conflicts of the watermark among clients according to the allocated keys by smart contracts. 
The client can achieve this goal by optimizing the loss function
by adding a regularization term into the objective functions, parameters will be trained to fit the watermark while fitting the main task.  In the above design, we embed the 0-1 watermark \textbf{K} into the positive and negative of the element of \textbf{AW}. 
\par And suppose the client choose the backdoor watermark which is derived from the backdoor attack. The way of choosing the trigger trained by the client in the edge distribution of the dataset or use a fixed texture directly. Here the fixed trigger is more suitable to act as the watermark for it can be stored on the blockchain permanently instead of depending on the training set to get the watermark. Through watermark negotiation, the client get the watermark, or in backdoor we call the trigger, \textbf{T}.  The client will mix up part of the training dataset \textbf{D} to get the backdoor dataset with simply $\alpha \textbf{D}+(1-\alpha) \textbf{T}$ and set their label as the target label. After constructing local dataset, the client will optimize the model with the local training objective of the normal task of cross entropy loss combined with the watermark mission $ \sum_{i=1}^{|B|} L\left(R\left(x_i, y_i\right) ; \omega\right)$ where R represent the backdoor watermark dataset.

There are many backdoor methods targeting at different kinds of models, the above takes the classification task as an example, but still, the backdoor method is not as universal as the feature method. By exploiting watermark injection process, the blockchain will record the mapping of the model and its unique watermark. The model can be stored and published on chain in the support of IPFS or other distributed storage technology. Anyone can query the chain to know the true own of the model which will be introduced soon, the on-chain process.

\subsubsection{On-chain Processing}
\par On the blockchain side, smart contracts complete tokenization by designing specific data structures to store the mapping of user addresses and watermarks. In the following discussion, we may use both the terms “token” and “watermark”, which refer to the same thing. Just as NFTs use metadata of digital collections as tokens, \textit{TokenizedModel} also uses the concept of watermarks to entitle models a storage-friendly and transferable non-fungible token. When the blockchain completes the watermark negotiation process with the client, a mapping between users and watermarks is established. The specific data structure of this mapping is shown in the right part of the Figure 2. When a user claims ownership of a model, the smart contract will obtain the model token owned by that user by searching this data structure and check whether the watermark can be detected in the model through a decentralized verification mechanism. For feature watermarks, we multiply the decryption matrix \textbf{A} and model parameters \textbf{W} to get the watermark; for backdoor watermarks, we mix trigger with the test data and pass into the model to observe whether the output is the agreed target class.
\par During the on-chain processing process, management and verification over model ownership relies on watermark embedding and recognition at the underlying algorithms which are introduced in local watermark injection. Below we will introduce how the platform provides users with various services based on these copyright verification function and try our best to shield the shortcomings of the underlying algorithm to improve the overall system’s verification reliability.

\subsection{Core Function of the Platform}
\par We list the services providing by \textit{TokenizedModel} in the yellow block of Figure 2. The core function of the entire framework is the recording and verification of model tokens. Various functions such as model transactions and model copyright sharing under federated learning scenario can be provided. We will first discuss the recording and verification of model tokens in further detail and then give the description of the rest two functions.
\subsubsection{Recording and Verification of Tokens }
\par As mentioned earlier, we use a dictionary-like data structure as the underlying container for storing tokens. The overall functionality of the system is heavily dependent on the accuracy of token verification. In order to achieve a more accuracy verification, apart from relying on further research in robust methods to embed model watermarking, our platform also provides certain mechanisms to shield as much as possible the shortcomings of the underlying algorithms. 
\par One of these mechanisms is the watermark negotiation process mentioned earlier. If the client chooses the watermark themselves and informs the blockchain, it will cause different client models to choose watermarks with a too close hamming distances, resulting in difficulty in discrimination. The watermark conflict will cause the difficulties in verification, Li \emph{et al.} has explored the solution in federated training process\cite{li2022fedipr}. In the unified platform, the watermark should be determined by the chain-side to avoid the conflict and bear some errors in watermark verification. Suppose we want to insert \textit{n} bits watermark by using feature-based method and the tolerance of bit error rate is \textit{r}, this means we still confirm the copyright if the different bits between the original watermark and  extracted watermark is less than $rn$. 
Hamming code indicate the counts of different bits between two bit streams when designing an error correcting code. And under this circumstance, the hamming distance of the original watermark of different client must satisfy a minimum distance of $2rn$.
Otherwise, the verification process will get confused due to a possibility of generating the same decoded result from two watermarks in the range of error tolerance and generate the conflict of copyright. The negotiation procedure is actually waiting the smart contracts to pick up a watermark that will not cause conflict, which cannot be done if we let the client to choose the watermark themselves.
\par For backdoor methods, the situation is similar with the feature-based method,  the smart contract tries to choose patterns with large differences for different clients, which lower the possible of training a close trigger and the upcoming copyright conflict. 

\par The second is the distributed verification process. Since the success rate of watermark verification is difficult to reach 100\% in real situations (this will be discussed in later case studies) and some malicious node will report a fake result during the verification process, we choose multiple distributed nodes to jointly perform the process of model verification, and the selection of distributed trusted nodes can adopt mechanisms such as committee mechanisms according to this article \cite{li2020blockchain}.
\par Finally, since the watermark information of the model will not be leaked to third parties, everyone only needs to trust the verification results given by the blockchain, avoiding various attacks on copyright due to watermark leakage.

\subsubsection{Model Ownership Transactions}
\par Just like the digital arts generate their value after the recording and ownership verification of NFT. Before we proposed the concept of \textit{TokenizedModel} which provides a platform to conveniently store their copyright information, once a model is made public online, it can be freely copied and its owner can hardly claim ownership of its work. It is also difficult to profit from the copies due to the lackness of the copyright. In this scenario where copying manner is unrestricted and there is no copyright protection, model transactions are rarely mentioned. After all, one can rarely tend to pay for a products that can be easily copied with no proof of ownership. However, tokenizing a model means that we entitles the model an attribute that can be circulated like an NFT. With digital currency assets on the blockchain, parties who agree to trade can simply modify the mapping relationship in the storage structure through smart contracts to complete model ownership transactions. And the blockchain will record the new owner forever and verify the ownership whenever. 
\par With the introducing of the concept of model transaction, the utilization of models can be more regulated and bring the model owner more profit than just providing the service on the internet facing the dangerous of stealing and unauthorized re-distribution.

\subsubsection{Model Copyright Sharing in Federated Learning} 
\par Federated learning is a common joint learning paradigm where multiple clients collaborate to complete training tasks for global models. Because the requirement for huge amount of dataset to train a well-performed model, individuals and small-scale company can hardly train a model on their own. Thus, federated learning is becoming a common way to train a model. At this time, it is necessary to record multiple watermarks and contributions (the recording structure can be seen in Figure 2) in the same model. When subsequent global models are used to provide services to others and generate revenue, incentives can be allocated to each client according to their contribution ratio to the model. In addition, during model transactions, a majority of “shareholders” must agree before a model ownership transaction can be executed. The record of model copyright for federated learning actually forms a decentralized autonomous organization (DAO) \cite{wang2019decentralized} among participating parties in training, making processes such as model training and incentive allocation more secure and reasonable. 
\par In summary, the entire architecture use watermark to tokenize models and provided some platform-based mechanism to help achieve a more reliable recording and verification procedure. By using the blockchain and by tokenizing the model, we bring many advantages of blockchain systems in copyright verification, we can treat the model like NFT to permanently record its copyright and execute transaction easily through cryptocurrency. And for federated learning, we have formed DAO-like organization to protect the collective interest by training participants all.   

\section{Case Study and Verification}
\par In this section, we demonstrate the effectiveness of our proposed framework in practical application scenarios through three case studies. These case studies include watermark embedding and verification for a single model, watermark embedding and verification for multi-client scenarios in federated learning, and verification based on backdoor watermarking. In the experiments, users embed watermarks during local model training. For the feature watermarking based on layer parameters, the extracted watermark from the model is compared bitwise with the original watermark to calculate the accuracy of watermark verification. For the backdoor watermarking, the accuracy of the model's judgments on backdoor samples is used as the watermark verification accuracy. In all experimental settings, we utilize the AlexNet network architecture and set the learning rate to 0.01.

\subsection{Case 1: Single Client stores the model on chain to record the copyright.}
\par This is the most common application scenario where users independently train a deep learning model, record the model's copyright ownership on the blockchain, and perform copyright verification and transactions. We conducted experiments on the image classification datasets CIFAR10 and CIFAR100. We embedded a 256-bit binary string in the model as a feature watermark. Additionally, to assess the robustness of the framework against watermark removal attacks, we added varying degrees of Gaussian noise to the models after completing the watermark embedding process, simulating malicious user interference with the model.


  

\begin{figure}
\centering
\includegraphics[width=0.5\textwidth]{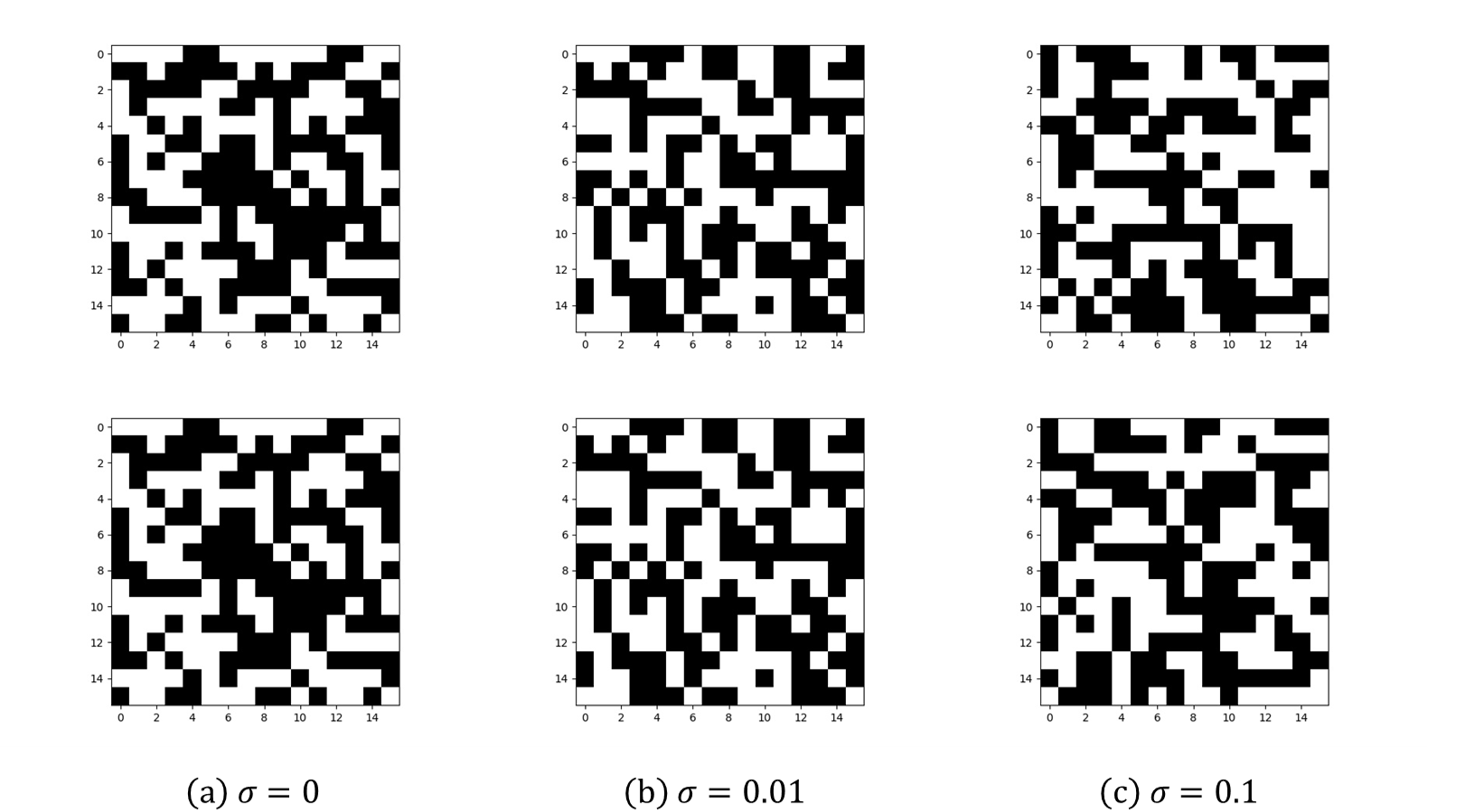}
 \caption{The original watermark (the first row) and the extracted watermark in the scene with different levels of Gaussian noise $\sigma$ (the second row) in CIFAR-10.}
  \label{fig:single-client-watermark-cifar10}
\end{figure}

  
\begin{figure}
\centering
\includegraphics[width=0.5\textwidth]{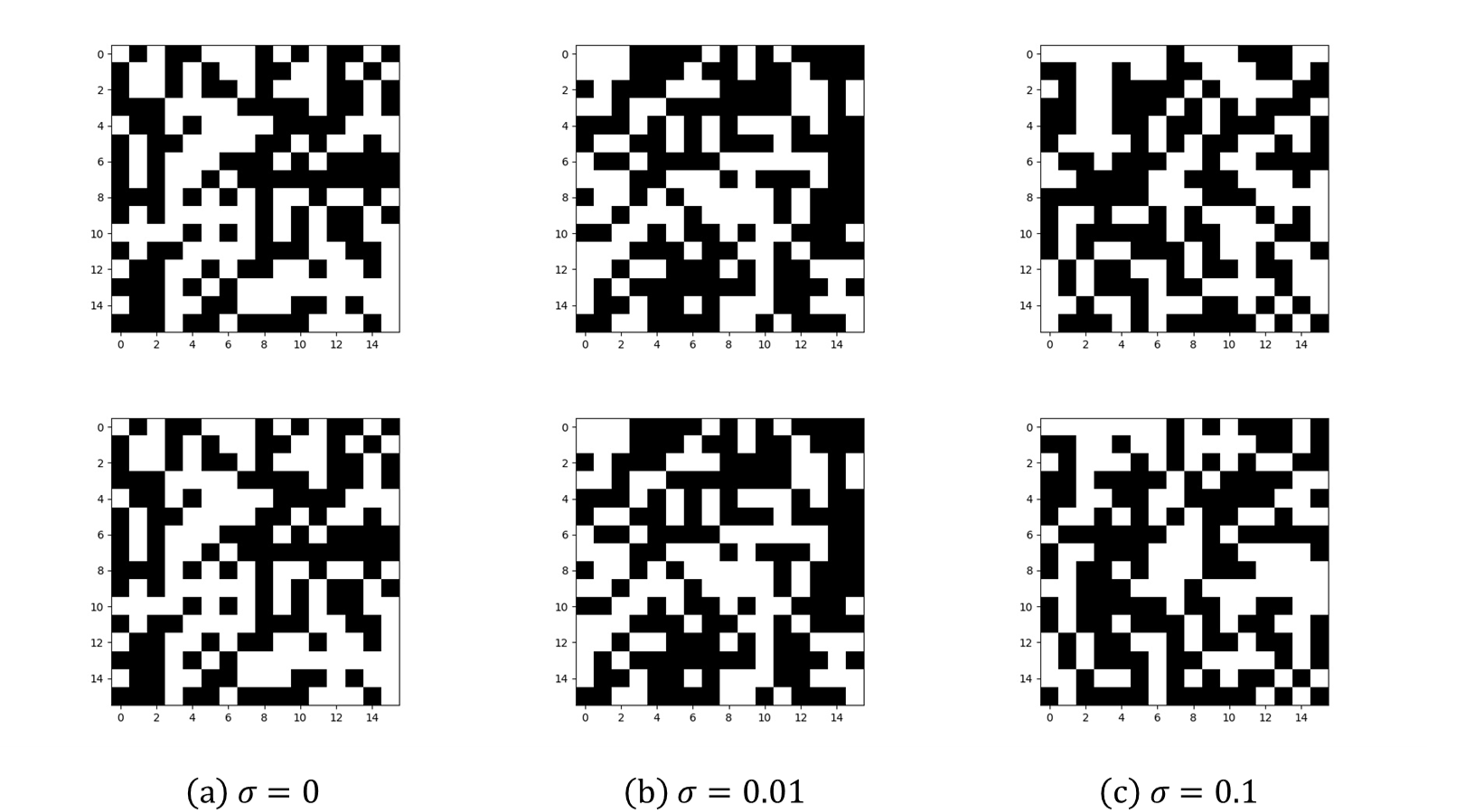}
 \caption{The original watermark (the first row) and the extracted watermark in the scene with different levels of Gaussian noise $\sigma$ (the second row) in CIFAR-100.}
  \label{fig:single-client-watermark-cifar100}
\end{figure}
A watermark verification accuracy of 100\% can be achieved in the absence of interference or low noisy interference. The addition of additional random noise has some impact on watermark extraction, but a high level of accuracy can still be maintained. When the Gaussian noise $\sigma$ reaches 0.1, the accuracy of both two datasets decreases to 0.890. Furthermore, the visual results of watermark extraction can be more intuitively observed from Figure \ref{fig:single-client-watermark-cifar10} and Figure \ref{fig:single-client-watermark-cifar100}. Hence, we can conclude that the watermark verification accuracy based on feature watermark has minimal correlation with the downstream task, as the watermark is directly injected into a specific layer of the model that contains redundant parameters to accommodate the watermark.

\subsection{Case 2: Federated Learning: The Global Model trained by multiple participant to record each copyright information of the contributor.}

\par This case is applicable to the scenario of federated learning, where a global model is jointly trained by multiple clients. In this case, the model does not belong to a single user but is collectively owned by multiple users. In practical applications, if the globally trained model in multi-client settings generates revenue, it becomes necessary to track the users involved in model training. This implies that multiple watermarks need to be embedded in the global model of federated learning, and the framework should have the capability of multi-watermark verification. In this case, we set up 10 clients, each with a network architecture and dataset identical to Case 1. Due to a higher probability of collusion, the size of the feature watermark is reduced to a smaller size to maintain the accuracy of verification.

\begin{figure}[htbp]
  \centering
  \begin{subfigure}[b]{0.24\textwidth}
    \centering
    \includegraphics[width=\textwidth]{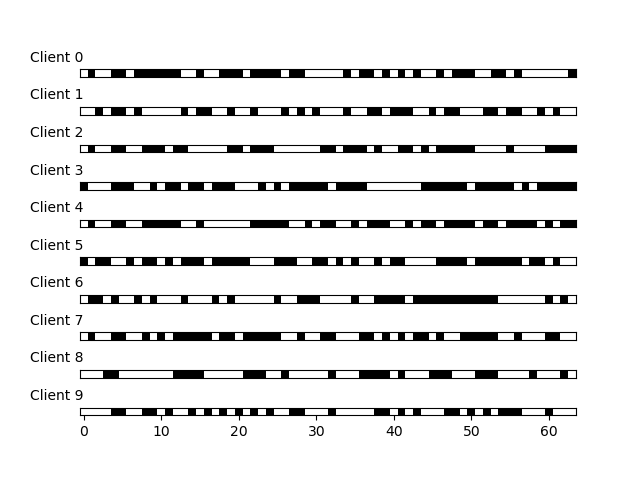}
    \label{fig:image5}
  \end{subfigure}%
  \hspace{-0.1cm}
  \begin{subfigure}[b]{0.24\textwidth}
    \centering
    \includegraphics[width=\textwidth]{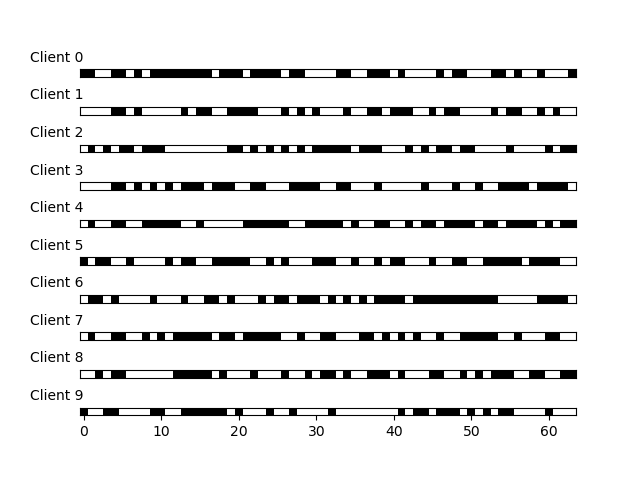}
    \label{fig:image6}
  \end{subfigure}
  
  \caption{The original watermark (left) and the extracted watermark (right) in federated learning in CIFAR-10.}
  \label{fig:multi-client-watermark-cifar10}
\end{figure}

\begin{figure}[htbp]
  \centering
  \begin{subfigure}[b]{0.24\textwidth}
    \centering
    \includegraphics[width=\textwidth]{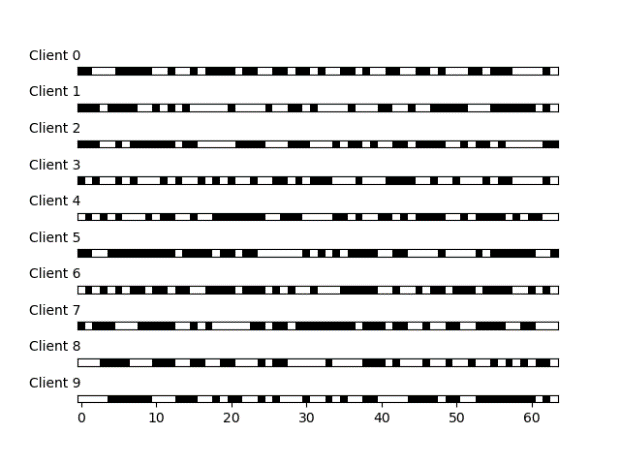}
    \label{fig:image5}
  \end{subfigure}%
  \hspace{-0.1cm}
  \begin{subfigure}[b]{0.24\textwidth}
    \centering
    \includegraphics[width=\textwidth]{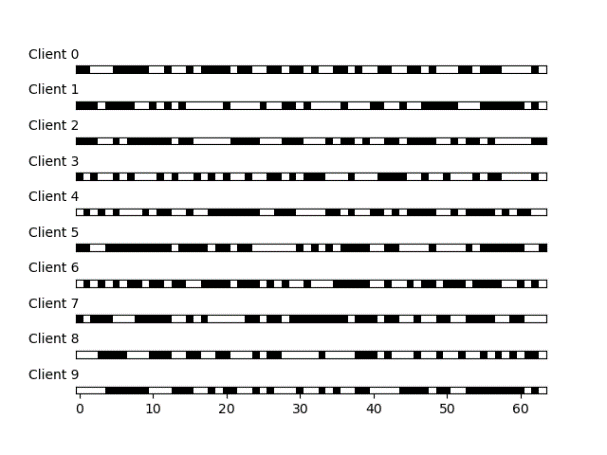}
    \label{fig:image6}
  \end{subfigure}
  
  \caption{The original watermark (left) and the extracted watermark (right) in federated learning in CIFAR-100.}
  \label{fig:multi-client-watermark-cifar100}
\end{figure}

\par Embedding and detecting multiple watermarks in the federated learning scenario poses relative difficulties. However, the watermark verification accuracy still remains around 80\%. Figure \ref{fig:multi-client-watermark-cifar10} and Figure \ref{fig:multi-client-watermark-cifar100} presents the visual results of the original watermarks and extracted watermarks for the multiple clients.

\subsection{Case 3:  Study of Backdoor Watermark.}

\par The backdoor-based watermark is distinct from the feature-based watermark as it relies on the model's outputs for specific backdoor samples to determine if the model contains a particular watermark. The same approach applies to the backdoor-based watermark. We employ the mixup method to obtain our own watermark dataset for injecting the backdoor watermark, which establishes a stronger relationship with the downstream task. Consequently, we can directly record the mixup image data into the blockchain like the data structure shown in Figure 2. The blockchain may have limited access to the detailed information regarding the downstream task of a specific client.

We demonstrates the verification accuracy using backdoor watermarking. For CIFAR-10 dataset, the single client accuracy is 0.958 and the accuracy for federated scenario is 0.847. 

\par By studying this case, we can find a good performance in both single-client and federated learning scenarios. For the single-client scenario, the watermark can be embedded without decreasing the main classification task. For federated scenario, the accuracy is lower than single-client due to the average aggregation strategy. Under this circumstance, the watermark embedding task from other client act also as the noisy and cause the decrease of accuracy. How to get a well embedding performance under the federation scenario is also a challenging problem.

\section{Conclusion and Future Work}

The overwhelming emergency of AIGC has drawn a lot of attention and also bring many copyright-concerned problems. In this work, a blockchain-empowered ownership verification platform was proposed to help regulate the management of the copyright on model and also entitled the model with a lot of financial advantages like NFTs. The proposed on-chain storage structure and decentralized verification stage help build the system more robust. 
\par However, works on model watermark should be continuously studied to provide more reliable bottom technology. Designing a watermark suitable for the generative model is an important content. Also for the federated scenario, how to make the watermark from multiparty can be embedded into a single model and how to design a mechanism to decrease the watermark conflict are also key contents towards a practical copyright platform. 



\bibliographystyle{IEEEtran}
\bibliography{IEEEabrv,reference}


\vspace{-10mm}
\begin{IEEEbiographynophoto}{Yihao Li}
received the B.S. degree in information engineering in 2022 from Sun Yat-sen University, Guangzhou, China, where he is currently working toward the Graduate degree in computer technology. His research focuses on the federated learning.
\end{IEEEbiographynophoto}
\vspace{-10mm}
\begin{IEEEbiographynophoto}{Yanyi Lai}
received the B.S. degree in software engineering in 2023 from South China Normal University, Guangzhou, China. He is currently working toward the Graduate degree in computer technology in Sun-Yat sen University. His research focuses on the federated learning.
\end{IEEEbiographynophoto}
\vspace{-10mm}
\begin{IEEEbiographynophoto}{Tianchi Liao}
is currently working toward the master’s degree with
the School of Computer Science and Engineering,
Sun Yat-Sen University, Guangzhou, China. Her
research interests include machine learning and
federated learning.
\end{IEEEbiographynophoto}
\vspace{-10mm}
\begin{IEEEbiographynophoto}{Chuan Chen}
received his Ph.D. degree from Hong Kong Baptist University, Hong Kong, in 2016. He is currently an Associate Professor at the School of Computer Science and Engineering, Sun Yat-sen University, Guangzhou, China. He has authored or co-authored over 80 international journal and conference papers. His research interests include federated learning, robust machine learning, and graph neural networks. He has also served as the Associate Editor for the journal Software Impacts.
\end{IEEEbiographynophoto}
\vspace{-10mm}
\begin{IEEEbiographynophoto}{Zibin Zheng}
received
the Ph.D. degree from the Chinese University of
Hong Kong, Hong Kong, in 2011. He is currently a
Professor with School of Data and Computer Science, Sun Yat-sen University, Guangzhou, China.
He is also the Chairman of the Software Engineering Department. He has authored or coauthored
more than 120 international journal and conference
papers, including three ESI highly cited papers.
According to Google Scholar, his papers have more
than 7000 citations, with an H-index of 42. His
research interests include blockchain, services computing, software engineering, and financial Big Data. He was the recipient of several awards, including
the Top 50 Influential Papers in Blockchain of 2018, ACM SIGSOFT Distinguished Paper Award at ICSE2010, Best Student Paper Award at ICWS2010.
He was BlockSys’19 and CollaborateCom’16 General Co-Chair, SC2’19,
ICIOT’18 and IoV’14 PC Co-Chair.
\end{IEEEbiographynophoto}
\vspace{-10mm}


\end{document}